\documentclass[aps,pra,twocolumn,groupedaddress,superscriptaddress]{revtex4-1}
\usepackage{bm}
\usepackage{amsmath}
\usepackage{graphicx}

\begin{document}


\title{Sympathetic Cooling of Mixed Species Two-Ion Crystals for Precision Spectroscopy}


\author{Jannes B. W\"ubbena}
\author{Sana Amairi}
\author{Olaf Mandel}
\affiliation{QUEST Institute for Experimental Quantum Metrology, Physikalisch-Technische Bundesanstalt, 38116 Braunschweig, Germany}
\author{Piet O. Schmidt}
\affiliation{QUEST Institute for Experimental Quantum Metrology, Physikalisch-Technische Bundesanstalt, 38116 Braunschweig, Germany}
\affiliation{Institut f\"ur Quantenoptik, Leibniz Universit\"at Hannover, 30167 Hannover, Germany}


\date{\today}

\begin{abstract}
Sympathetic cooling of trapped ions has become an indispensable tool for quantum information processing and precision spectroscopy. In the simplest situation a single Doppler-cooled ion sympathetically cools another ion which typically has a different mass. We analytically investigate the effect of the mass ratio of such an ion crystal on the achievable temperature limit in the presence of external heating. As an example, we show that cooling of a single Al$^+$ with Be$^+$, Mg$^+$ and  Ca$^+$ ions provides similar results for heating rates typically observed in ion traps, whereas cooling ions with a larger mass perform worse. Furthermore, we present numerical simulation results of the rethermalisation dynamics after a background gas collision for the Al$^+$/Ca$^+$ crystal for different cooling laser configurations.
\end{abstract}


\maketitle

\section{Introduction}
Sympathetic laser cooling of trapped ions is an important experimental tool in diverse fields, such as quantum information processing, chemistry and precision spectroscopy. In quantum information processing, a sympathetic cooling ion can be used to cool the qubit ions in a multiplexed trap architecture \cite{kielpinski_architecture_2002}, without affecting the quantum information store in the internal states of the qubit ions \cite{home_memory_2009, home_complete_2009}. Sympathetic cooling of atomic and molecular ions enables studies of chemical reactions at cold and ultra-cold temperatures \cite{willitsch_chemical_2008, gingell_cold_2010, staanum_probing_2008}. A major interest in sympathetic cooling of clouds of ions stems from spectroscopy of atomic \cite{imajo_high-resolution_1996, roth_sympathetic_2005, rosenband_observation_2007} and molecular ion species \cite{mlhave_formation_2000, blythe_production_2005, roth_production_2006, roth_rovibrational_2006} with a complex internal level structure that can not be laser cooled directly. Precision spectroscopy of single or few spectroscopy ions sympathetically cooled by a well-controllable cooling or logic ion in a linear ion trap can be performed using quantum logic spectroscopy (QLS) \cite{wineland_quantum_2002, schmidt_spectroscopy_2005}. Here, the laser-cooled logic ion not only provides sympathetical cooling, but also assists in the readout process of the spectroscopy ion after interrogation. After its successful implementation in optical frequency standards based on aluminium ions \cite{Rosenband2008,Chou2010}, several experiments including spectroscopy of molecular ions \cite{schiller_tests_2005, kajita_estimated_2011}, highly-charged ions \cite{wineland_experimental_1998, gruber_evidence_2001, schiller_hydrogenlike_2007}, super-heavy ions \cite{drewsen_cooling_2007}, and metal ions \cite{Hemmerling2011} using this technique have been proposed.
One of the requirements for these applications is efficient Doppler laser cooling of small linear ion crystals, which in its simplest form consists of two ions of (in general) unequal mass. Doppler laser cooling has first been experimentally demonstrated with trapped magnesium ions \cite{wineland_radiation-pressure_1978}. It was soon realized that the strong mutual electro-static interaction between a laser cooled ion species and another species not interacting with the cooling laser, allowed sympathetic cooling of the latter in large clouds of ions \cite{drullinger_high-resolution_1980, larson_sympathetic_1986}. Crystallized linear chains of up to 15 ions have been sympathetically cooled with a single cooling ion of similar mass \cite{bowe_sympathetic_1999, rohde_sympathetic_2001, blinov_sympathetic_2002}. Sympathetic Doppler and even ground-state cooling of two-ion crystals with a mass ratio of up to three has been successfully implemented \cite{barrett_sympathetic_2003, schmidt_spectroscopy_2005}. 
Cooling of even larger mass ratios has been proposed by trapping the two species in separate potential wells \cite{heinzen_quantum-limited_1990,hasegawa_sympathetic_2011}. The structure, dynamics and cooling of linear ion crystals composed of ions with unequal mass have been investigated theoretically mostly in the context of quantum information processing \cite{alekseev_sympathetic_1995, james_quantum_1998, kielpinski_sympathetic_2000, morigi_two-species_2001, hasegawa_limiting_2003}, with particular emphasis on the mode structure and the cooling rates. For applications, such as optical frequency standards requiring high spectroscopic accuracy, the lowest achievable temperature during Doppler cooling will determine relativistic and trapping-field induced shifts \cite{rosenband_frequency_2008, Chou2010}. Motional heating of the ions in the trap due to electric field fluctuations \cite{wineland_experimental_1998, leibrandt_modeling_2007, safavi-naini_microscopic_2011} significantly modifies the achievable Doppler cooling temperature.

Here, we develop an analytical model to study the achievable motional energy using sympathetic cooling in linear two-species two-ion crystals within the pseudopotential approximation in the presence of external motional heating. The achievable cooling limit strongly depends on the strength of the external electric field fluctuations, the mass ratio between cooling and spectroscopy ions, and the trap parameters. We show that in particular, the Doppler cooling temperature in radial direction is much more sensitive to a mass mismatch compared to the axial direction. We use this model to investigate the second-order Doppler shift for Doppler cooled ion crystals in an ${}^{27}$Al$^+$ quantum logic clock with different logic ion species. We show that next to the obvious choice of ${}^{25}$Mg$^+$-ions which excel because of their almost perfect mass match, ${}^{40}$Ca$^+$-ions will perform similarly and in some situations even better than the Mg ions, owing to the lower achievable Doppler cooling temperature. Besides the cooling limit, the required cooling time after e.g. a collision with the hot background gas is an important aspect. We numerically simulate the cooling dynamics of a Ca$^+$/Al$^+$ ion crystal after such a collision event. Before crystallization of the two-ion crystal, the cooling rate is comparable to the single ion case. After crystallization, the motion of the ions is described in normal modes and the cooling rate is limited by modes that are only weakly cooled by the logic ion.

In section \ref{Modes} we briefly recall the low temperature dynamics of the two-ion crystal and introduce analytic formulae for the six mode frequencies and the modal amplitudes. In section \ref{CoolingLimits} we expand the standard Doppler cooling model to the two-ion crystal case and investigate the effect of external heating on the temperature limit. In section \ref{CoolingTimes} we describe our cooling dynamics simulation and discuss the results before summarizing in section \ref{Summary}.

\section{Normal Modes of a Two-Ion Crystal\label{Modes}}
\begin{figure}
\includegraphics{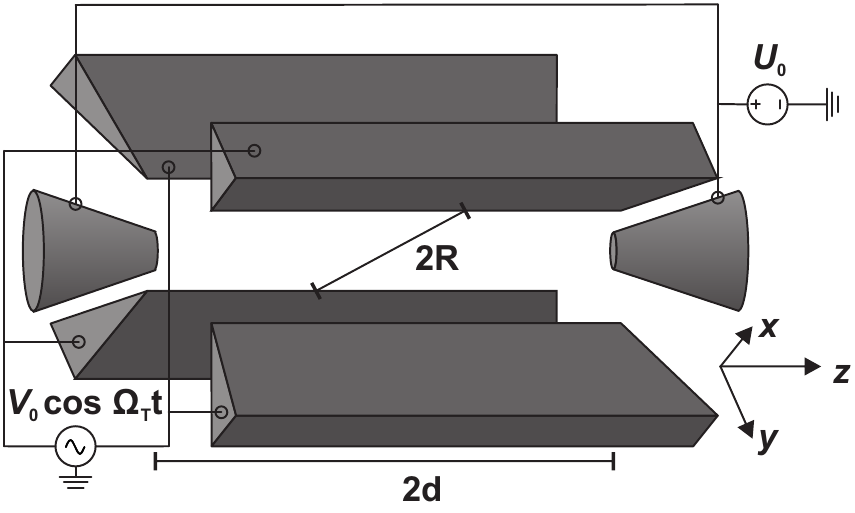}
\caption{Schematic of a linear ion trap. The trap consists of four blade-shaped electrodes of which two opposing ones are connected to an RF-voltage while the other two are connected to ground. It also includes two endcap electrodes that are connected to a positive DC voltage. The line through the two endcaps defines the trap axis $z$ and the other two axes ($x$, $y$) are chosen such that two blades lie on each axis.\label{IonTrap}}
\end{figure}
In linear Paul traps the confinement of charged particles is realised by two distinct electric fields \cite{ghosh_ion_1995}. The first is a rapidly oscillating 2D electric quadrupole in the radial ($xy$) plane and the second is a static 3D quadrupole field providing confinement along the trap ($z$) axis. While the first is assumed to be purely radial and have no field components in the axial direction, the latter must have components in the radial directions additionally to its axial field in order to satisfy Laplace's equation. Following the notation of \cite{Wineland1998} the total trap potential is given by
\begin{multline}
\Phi(x,y,z,t) = \frac{V_0}{2} \cos \Omega_T t \left( 1 + \frac{x^2-y^2}{R'^2}\right) \\ 
 + \kappa U_0\left( \frac{z^2 - \alpha x^2 - (1-\alpha) y^2}{d^2}\right) \quad , \label{TotalPotential}
\end{multline}
where $V_0$ and $U_0$ are the applied RF and DC voltages, $R' \approx R$ and $d$ are the radial and axial trap dimensions, respectively, $\kappa$ is a geometric factor \cite{Raizen1992}, $0 < \alpha < 1$ is a parameter indicating the radial asymmetry of the static field, and $\Omega_T$ is the angular frequency of the rapidly oscillating RF-field (see Fig.~\ref{IonTrap}). In the pseudopotential approximation, a single ion of mass $m$ and charge $e$ in this trap will experience a potential of the form  \cite{paul_electromagnetic_1990}
\begin{equation}
 U(x,y,z)=\frac{1}{2}m\omega_x^2 x^2 + \frac{1}{2}m\omega_y^2 y^2 + \frac{1}{2}m \omega_z^2 z^2.
\end{equation}
In this potential, the $\omega_{x,y,z}$ denote the single ion trap frequencies, i.e. the frequencies of the ion's secular motion along the different axes. They are given by
\begin{eqnarray}
\omega_z &=& \sqrt{\frac{2 e \kappa U_0}{md^2}}\\
\omega_x &=& \sqrt{\omega_p^2-\alpha \omega_z^2}\\
\omega_y &=& \sqrt{\omega_p^2-(1-\alpha) \omega_z^2}
\end{eqnarray}
where $\omega_z$, $\omega_x$ and $\omega_y$ are the axial and two radial trap frequencies, respectively, and
\[ 
\omega_p = \frac{e V_0}{\sqrt{2} \Omega_T m R'^2}
\]
 describes the contribution of the RF potential to the radial trap frequencies \cite{Wineland1998}. Introducing a factor $\epsilon = \omega_p / \omega_z$ as in \cite{kielpinski_sympathetic_2000} simplifies the radial trap frequencies to 
\begin{equation}
 \omega_x = \sqrt{\epsilon^2 - \alpha} \, \omega_z, \qquad \omega_y = \sqrt{\epsilon^2 - (1-\alpha)} \, \omega_z.
\end{equation}

From the known trap frequencies, the $\alpha$ and $\epsilon$ parameters for a given ion with mass $m=m_1$ in a given trap can be derived. This allows the computation of frequencies for ions with different masses $m_2$ in the same trap. While the axial trap frequencies simply scale with the square root of the mass, the radial trap frequencies additionally depend on $\epsilon$ and $\alpha$: 
\begin{align}
 \omega_{z,2}&=\sqrt{\frac{m_1}{m_2}}\omega_{z,1} \\
 \omega_{x,2}&=\sqrt{\frac{m_1}{m_2}}\sqrt{\frac{\frac{m_1}{m_2}\epsilon^2-\alpha}{\epsilon^2-\alpha}}\omega_{x,1} \\
 \omega_{y,2}&=\sqrt{\frac{m_1}{m_2}}\sqrt{\frac{\frac{m_1}{m_2}\epsilon^2-(1-\alpha)}{\epsilon^2-(1-\alpha)}}\omega_{y,1}  
\end{align}
For the remainder of this work we assume $\alpha=1/2$ to simplify the algebra.

\begin{figure*}
\includegraphics{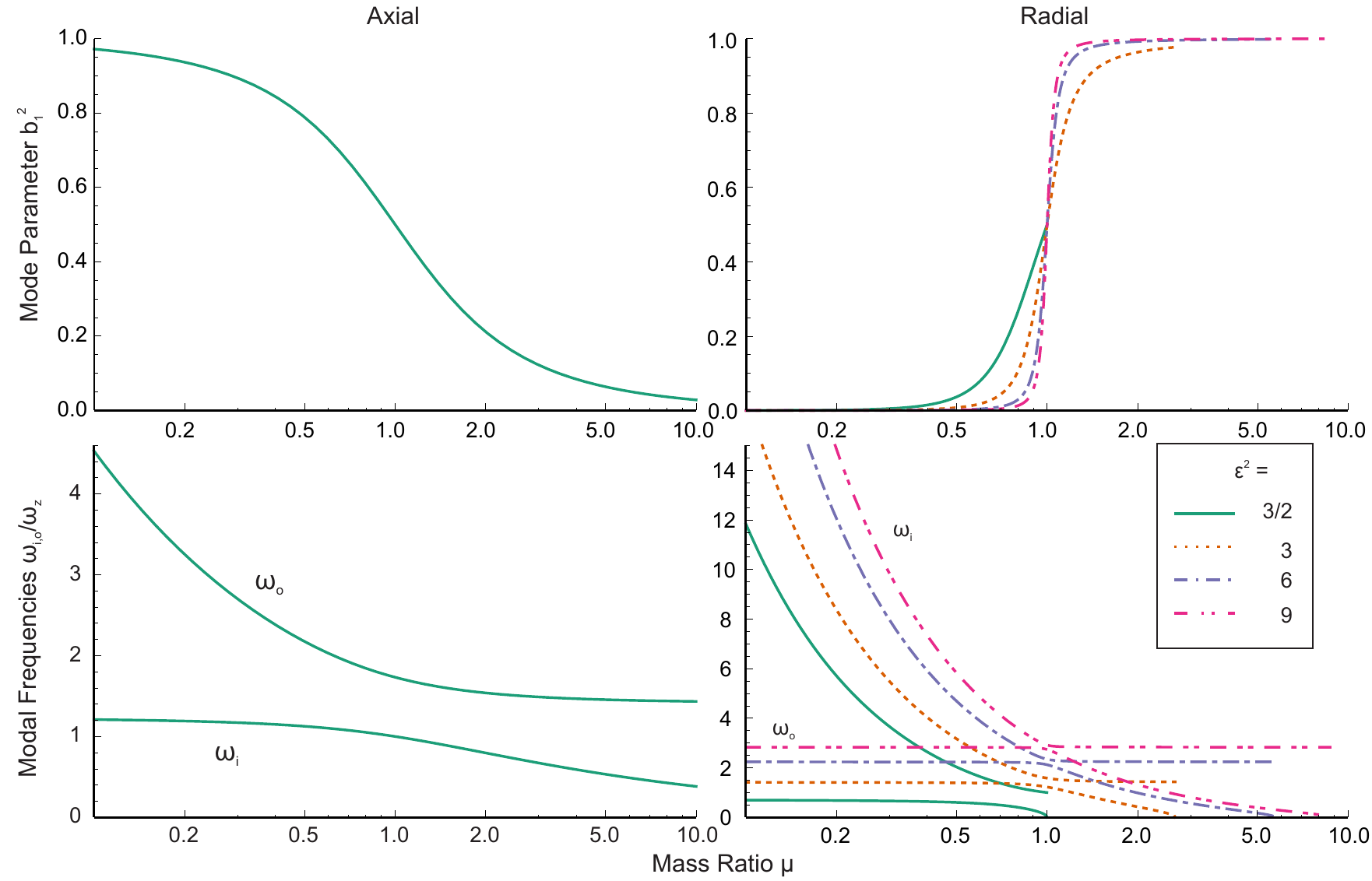}
\caption{Normal mode frequencies and normalized amplitudes for a two-ion two-species crystal. The square of the normalized amplitude ($b_1^2$) and the in-phase ($\omega_i$) and out-of-phase ($\omega_o$) trap frequencies normalized to the axial trap frequency of a single ion of mass $m_1$ are shown for the axial and radial direction for different mass ratios $\mu=m_2/m_1$ and $\epsilon$-parameters.\label{FrequenciesTwoIons}}
\end{figure*}

If two ions are simultaneously trapped  in the same linear Paul trap and strongly cooled to near 0~K temperature, they will eventually crystallize at equilibrium positions along the trap axis, equally spaced at a distance $z_0 = \left( \frac{e d^2}{32 U_0 \pi \epsilon_0}\right)^{1/3}$ from the trap centre \cite{wineland_experimental_1998, james_quantum_1998}. The remaining motion of ions 1 and 2 can then be described as small, coupled oscillations $q_1, q_2$ around these equilibrium positions. Along every principal axis the motion consists of a superposition of an out-of-phase mode ($o$) where the two ions always move in opposite directions and an in-phase mode ($i$) where the two ions move in the same direction. Following the approach of \cite{kielpinski_sympathetic_2000}, the oscillations along a chosen direction are given by
\begin{align}
q_1(t) &= z_i b_1 \sin (\omega_i t + \phi_i) + z_o b_2 \cos (\omega_o t + \phi_o) \label{IonMotion1}\\ 
q_2(t) &= \frac{z_i b_2}{\sqrt{\mu}} \sin(\omega_i t + \phi_i) - \frac{z_o b_1}{\sqrt{\mu}} \cos(\omega_o t + \phi_o) \label{IonMotion2}
\end{align}
where $\omega_{i,o}, \phi_{i,o}$ are the angular eigenfrequencies and phases of the in-phase and out-of-phase modes, respectively, and $b_{1,2}$ are the components of the normalised eigenvector of the in-phase mode, satisfying $b_1^2+b_2^2=1$, in a coordinate system where the motion of the second ion is scaled by a factor of $1/\!\sqrt{\mu}$ with $\mu=m_2/m_1$. The $z_{i,o}$ are the modal amplitudes (see Eq.~(\ref{EnergyI}),~(\ref{EnergyO})). The calculation of the modal frequencies and the $b_{1,2}$-parameters can be performed similarily to for example \cite{kielpinski_sympathetic_2000}: For every ion the sum of the trap pseudopotential and the Coulomb potential due to repulsion from the other ion is developed around the equilibrium positions and the coupled equations of motions are solved in lowest order, neglecting higher order non-linear couplings \cite{wineland_experimental_1998, nie_theory_2009}. The results of this calculation for ions with different mass ratios $\mu$ and different $\epsilon$-parameters are given by:

\begin{align}
 \omega_{i,z} &= \sqrt{\frac{1 + \mu - \sqrt{1-\mu+\mu^2}}{\mu} } \omega_z\\
 \omega_{o,z} &= \sqrt{\frac{1 + \mu + \sqrt{1-\mu+\mu^2}}{\mu} } \omega_z\\
 b_{1,z}^2 &= \frac{1-\mu+\sqrt{1-\mu+\mu^2}}{2 \sqrt{1-\mu+\mu^2}} \\
\omega_{i,x,y} &= \sqrt{-\frac{\mu +\mu ^2-\epsilon ^2 \left(1+\mu ^2\right)-a}{2 \mu ^2}} \omega_z\\
 \omega_{o,x,y} &= \sqrt{-\frac{\mu +\mu ^2-\epsilon ^2 \left(1+\mu ^2\right)+a}{2 \mu ^2}} \omega_z\\
 b_{1,x,y}^2 &= \frac{\mu -\mu ^2+\epsilon ^2 \left(-1+\mu ^2\right)+a}{2 a} \label{RadialBFactor}
\end{align}
where the parameter
\begin{equation}
 a = \sqrt{\epsilon ^4 \! \left(\mu ^2 \!\! - \!\! 1\right)^2 \!\! -\!\! 2 \epsilon ^2 (\mu \!\! - \!\! 1)^2 \mu  (1 \!\! + \!\! \mu ) \!\! + \!\! \mu ^2 (1 \!\! + \!\! (\mu \!\! - \!\! 1) \mu )}
\end{equation}
was introduced. The $b_2$ parameters calculate as $b_2 = \sqrt{1-b_1^2}$.  Figure \ref{FrequenciesTwoIons} shows the calculated eigenmode amplitudes and frequencies in axial and radial directions. The heavier of the two ions has the largest amplitude for the mode with the lowest frequency, which is the axial in-phase mode and radial out-of-phase mode. It is worthwhile noting that the radial mode amplitudes are much more sensitive to a change in the mass ratio. As a consequence, the radial motion of the ions is nearly decoupled for ion species with mass ratios $\mu < 0.25$ or $\mu > 4$ for typical traps, i.e. for one mode ion 1 has a large normal mode amplitude and ion 2 has a small one, whereas for the other mode the situation is reversed. The radial curves end at mass ratios where the radial out of phase mode frequencies reaches zero because for higher mass ratios the ion crystal turns from a linear axial configuration to a linear radial configuration \cite{rafac_stable_1991, schiffer_phase_1993}.

In this notation, the total energies of the two modes (i.e. kinetic + potential energy) along one direction (valid for radial and axial directions) are given by

\begin{align}
E_i &= \frac{1}{2} m_1 z_i^2 \omega_i^2 \qquad \text{and} \label{EnergyI}\\
E_o &= \frac{1}{2} m_1 z_o^2 \omega_o^2 \label{EnergyO} .
\end{align}

\section{Cooling limits\label{CoolingLimits}}
\subsection{Doppler cooling of an isolated system}
In the following, we will briefly outline the usual derivation of the Doppler cooling limit \cite{Lett1989,Wineland1979,Hänsch1975}, before expanding the model to take into account an additional heating rate in the next section. The model differs from the standard derivation in so far as it takes into account the modal structure of the two-ion crystal right from the beginning.

Doppler cooling of all modes is achieved by exposing ion 1 to laser radiation of intensity $I$ and an angular frequency $\omega = \omega_0  + \Delta$ detuned by $\Delta$ from ion 1's cooling transition with angular frequency $\omega_0$, FWHM linewidth $\Gamma$ and saturation intensity $I_0$. The $\boldsymbol{k}$-vector of the radiation is given by $\boldsymbol{k} = (l_x, l_y, l_z) k = (l_x, l_y, l_z) \omega_0/c$ with the unit vector $(l_x, l_y, l_z)$.

Every scattering event of the radiation with the cooling ion will on average change the momentum of the system by $\hbar \boldsymbol{k}$. Due to the red detuning ($\Delta<0$) of the cooling light this is more likely to happen when the cooling ion moves towards the laser such that on average energy is removed from the ion crystal. The average energy loss rate can be calculated by averaging the momentum change over the periods of both oscillatory modes in each direction. For sufficiently cold ions ($|l_x k \dot q_1| \ll \Gamma$) in the weak binding regime ($\Gamma \gg \omega_{i,o}$), these can (exemplarily in the x-direction) be shown to be (see Appendix \ref{RatesCalculation})
\begin{align}
\frac{dE_i}{dt}_{\text{cool}} &\approx \hbar l_x^2 k^2 \frac{I}{I_0} \frac{2 \Delta/\Gamma}{(1+I/I_0 + (2\Delta/\Gamma)^2)^2} z_i^2 b_1^2 \omega_i^2 \label{CoolingRateEquation}\\
\frac{dE_o}{dt}_{\text{cool}} &\approx \hbar l_x^2 k^2 \frac{I}{I_0} \frac{2 \Delta/\Gamma}{(1+I/I_0 + (2\Delta/\Gamma)^2)^2} z_o^2 b_2^2 \omega_o^2.
\end{align}

The competing heating rate (considering the statistical distribution of the momentum changes due to both the absorption and the spontaneous emission of the cooling photons) is given by (see Appendix \ref{RatesCalculation})
\begin{align}
\frac{dE_i}{dt}_{\text{heat}} &\approx \hbar^2 (3 l_x^2+ 1)k^2 \frac{I}{I_0} \frac{\Gamma}{12 m_1} \frac{b_1^2}{1+I/I_0 + (2\Delta/\Gamma)^2}\label{TotalHeatingRate}\\
\frac{dE_o}{dt}_{\text{heat}} &\approx \hbar^2 (3 l_x^2+ 1)k^2 \frac{I}{I_0} \frac{\Gamma}{12 m_1} \frac{b_2^2}{1+I/I_0 + (2\Delta/\Gamma)^2}.
\end{align}

The cooling rate is a function of the ion's scattering rate and proportional to the square of the k-vector component ($l_x^2$) of the cooling light along the considered mode direction, times the square of the motional amplitude of the cooling ion for that mode ($z_i^2b_1^2$, $z_o^2b_2^2$). In contrast, the heating rate has a component from the isotropic photon emission in addition to directed absorption from the cooling laser beam.

The steady state solution is obtained from a balance between cooling and heating rates
\begin{equation} \label{eq:steadystate}
\frac{dE_{i,o}}{dt}_{\text{heat}} + \frac{dE_{i,o}}{dt}_{\text{cool}} = 0.
\end{equation}
The resulting cooling limit can be expressed as
\begin{equation}
E_\mathrm{limit} = \frac{\hbar(4 \Delta^2 + \Gamma^2 (1+I/I_0))(1+3 l_x^2)}{48 |\Delta| l_x^2} \label{CoolingLimitEquation}
\end{equation}
for both modes. The cooling limit is independent of the mode eigenvectors $b_{1,2}$ since the heating and cooling processes act in the same way on the motion of the ion crystal. As a consequence, modes with a small eigenvector component experience cooling and heating rates that are reduced by the same amount and will limit the time it takes to reach the steady-state temperature (see section \ref{CoolingTimes}). In the case of very low cooling intensity ($I/I_0 \rightarrow 0$), optimum detuning ($\Delta = - \Gamma/2$) and a laser that cools all directions symmetrically ($l_x = l_y = l_z = 1/\sqrt{3}$), this results in the well known Doppler cooling limit
\begin{equation}\label{eq:DClimit}
E_D = \hbar \Gamma / 2 .
\end{equation}

Very low cooling intensity leads to the lowest theoretical cooling limit at the expense of long cooling times, since the cooling rate gets very small. In practice, cooling intensities close to the saturation intensity are typically used. A larger cooling rate renders the system more robust if exposed to additional external heating. The maximum cooling rate is achieved at $I = 2 I_0$ and $\Delta = -\Gamma/2$. However, this choice of parameters increases the cooling limit by a factor 2.

\begin{figure}
\includegraphics{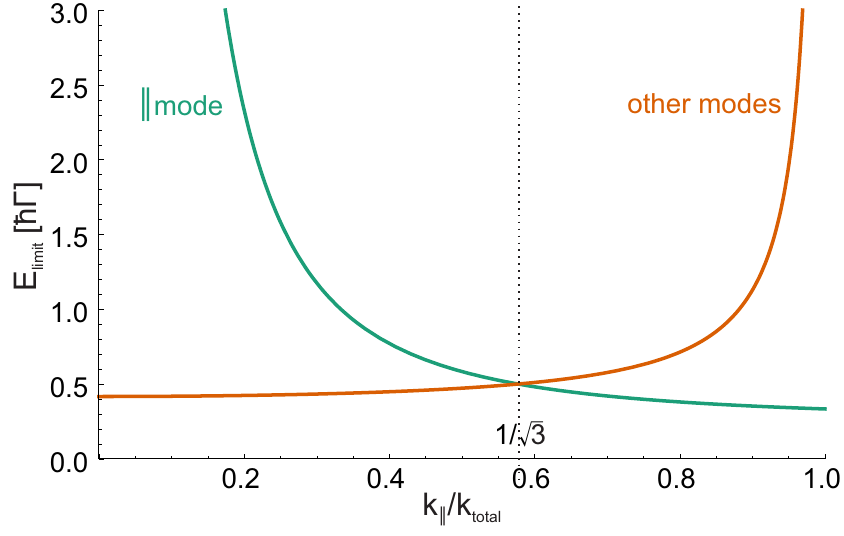}
\caption{Dependence of the cooling limits on the angle of the cooling laser. $k_\parallel / k_{\text{total}}$ denotes the component of the laser radiation parallel to the investigated axis. The other two axes are assumed to be cooled equally.\label{DependencyOnLaserAngle}}
\end{figure}

Aligning the cooling beam along the direction of a particular set of modes, cooling below the Doppler-cooling limit given by Eq.~(\ref{eq:DClimit}) in this direction is possible. However, the cooling limit in the other directions will be strongly increased as can be seen in Fig.~\ref{DependencyOnLaserAngle} and has been discussed in \cite{javanainen_light-pressure_1980}. This is a direct consequence of the cooling rate of a specific mode being dependent only on the k-vector projection along this direction, whereas the heating rate has a contribution from the isotropic spontaneous emission heating.

Precision spectroscopy and atomic frequency standards based on trapped ions require small kinetic energies to achieve small second-order Doppler shifts, which is particularly important for light ions such as $^{27}$Al$^+$. This shift is given by \cite{Riehle2005}
\begin{equation}
\frac{\Delta \nu}{\nu} = - \frac{\langle v^2 \rangle}{2 c^2},
\end{equation}
where $\langle v^2 \rangle$ is the average value of the square of the ion's velocity and $c$ is the speed of light. This relation allows a direct mapping between cooling limits and corresponding relativistic Doppler shifts. In the absence of external heating rates, both ions will have the same energy given by Eq.~(\ref{eq:DClimit}) and the second-order Doppler shift of the clock ion is
\begin{equation}
\frac{\Delta \nu}{\nu} = -\frac{\hbar \Gamma}{4 m_2 c^2}.
\end{equation}
In the case of ${}^{27}$Al$^+$ sympathetically cooled by ${}^{40}$Ca$^+$, this results in shifts of $9.2 \times 10^{-19}$ for each of the three directions. For the radial modes, this number has to be multiplied by a factor of roughly 2 (see Appendix \ref{AppendixMicromotion}) to account for the kinetic energy in the intrinsic micromotion of the radial motion of the clock ion which follows from generalisation of the results of \cite{Berkeland1998}. The only influence the selected cooling ion species has on this result is the linewidth of its cooling transition. Furthermore, the mass ratio determines the time it takes to reach steady state, but has no influence on the cooling limit. This changes as soon as external heating rates are included in the model.

\subsection{Doppler cooling with external heating rates} \label{ExternalHeatingRates}
External heating rates of cooled ion crystals are assumed to be mostly due to stochastic electric field fluctuations  \cite{Wineland1998,turchette_heating_2000,Deslauriers2006, safavi-naini_microscopic_2011}. If the characteristic distance between the electrodes and the ions is much larger than the distance between the ions in the crystal, the electric field across the ion crystal can be assumed constant and the heating rate due to field fluctuations can be written as \cite{kielpinski_sympathetic_2000}:
\begin{align}
\frac{dE_i}{dt}_{\text{fluct}} &= \frac{q^2 S_E}{4 m_1}\left(b_1 + \frac{1}{\sqrt{\mu}} b_2 \right)^2 \label{AddHeating1}\\
\frac{dE_o}{dt}_{\text{fluct}} &= \frac{q^2 S_E}{4 m_1}\left(b_2 - \frac{1}{\sqrt{\mu}} b_1 \right)^2 .\label{AddHeating2}
\end{align}

Here $S_E=S_E(\omega)$ denotes the electric field spectral density, which is assumed to be spectrally constant for the relevant $\omega_{i,o}$. The homogeneous field fluctuations will only couple to centre of mass motion and therefore much more strongly to the in-phase mode than the out-of-phase mode. In fact for $\mu=1$ it follows that $b_1 = b_2$ and the out-of-phase mode is not heated at all. The total energy injected in one direction of motion is obtained by adding the in-phase and out-of-phase heating rates, which turns out to be proportional to $m_1^{-1}+m_2^{-1}$, showing the advantage of heavy cooling and clock ions.

The cooling limit in the presence of external heating is obtained by including the heating rates Eq.~(\ref{AddHeating1}) and Eq.~(\ref{AddHeating2}) in the steady state condition Eq.~(\ref{eq:steadystate}). The steady state energy limit for the in-phase mode (substitute $b_1$ by $b_2$ and $b_2$ by $-b_1$ for the out-of-phase-mode) is given by:

\begin{multline}
E_{\text{limit},i} = \frac{\Gamma \left( 1+(2\Delta/\Gamma)^2+I/I_0\right)^2}{48 |\Delta| \hbar I/I_0 l_x^2 k^2 } \times \\
\left[ \frac{\Gamma \hbar^2 I/I_0 (3l_x^2 + 1)k^2}{1 + (2 \Delta/\Gamma)^2 + I/I_0} + \frac{3}{b_1^2} (b_1+\frac{1}{\sqrt{\mu}}b_2)^2 q^2 S_E\right]
\label{reallimit}
\end{multline}
The additional heating modifies the steady state solution for Doppler cooling (Eq.~(\ref{CoolingLimitEquation})) by breaking the symmetry between cooling and heating processes: The photon-induced heating and cooling rates have the same dependence on the mode amplitudes, resulting in a Doppler limit independent of this parameter. The external heating rate has no cooling component and a more complex dependence on the mode amplitudes, resulting in a cooling limit that is a sensitive function of the modal amplitudes and therefore the $\epsilon$ and $\mu$ parameters. As a result, modes where the cooling ion has a large relative amplitude are cooled more efficiently than modes where it has a small amplitude.

The two terms in the square bracket in Eq.~(\ref{reallimit}) give the contribution of the photon and the external heating rate to the total heating rate, respectively. It is instructive to investigate the two extreme cases, in which either one of the contributions dominate. Defining an electric field spectral density $S_{E0} = \Gamma \hbar^2 k^2 / 12 q^2$ \cite{SE0_calcium} for which the two heating rates in a symmetric ($\mu=1$) ion crystal and symmetric cooling in all three directions ($l_x=1/\sqrt{3}$) equal $\frac{dE_i}{dt}_{\text{fluct}} = \frac{dE_i}{dt}_{\text{heat}}(\Delta = -\Gamma/2, I = 2 I_0)$, the two cases are

\begin{enumerate}
\item $S_E \ll S_{E0}$:
 In this case the photon heating rate at the point of maximum cooling is much larger than the external heating and the latter can therefore be neglected. By choosing the intensity $0 < I/I_0 < 2$ optimally, the cooling limit of Eq.~(\ref{CoolingLimitEquation}) is recovered:
\begin{equation}
 \hbar \Gamma / 2 < E_{\text{limit}} < \hbar \Gamma
\end{equation}
\item $S_E \gg S_{E0}$:
 In this regime the photon heating rate can be neglected and the cooling limit will proportionally depend upon the electric field spectral density and the mode amplitudes. Here the optimum intensity is that of maximum cooling rates ($I=2 I_0$).
\begin{align}
 E_{\text{limit},i} = \frac{q^2 S_E (b_1 + \frac{1}{\sqrt{\mu}}b_2)^2 \Gamma \left( 1+(2\Delta/\Gamma)^2+I/I_0\right)^2}{16 b_1^2 |\Delta| \hbar l_x^2 k^2 I/I_0 } \\
 E_{\text{limit},o} = \frac{q^2 S_E (b_2 - \frac{1}{\sqrt{\mu}}b_1)^2 \Gamma \left( 1+(2\Delta/\Gamma)^2+I/I_0\right)^2}{16 b_2^2 |\Delta| \hbar l_x^2 k^2 I/I_0 } 
\end{align}
\end{enumerate}

In both regimes the best cooling performance is always achieved at a detuning of $\Delta = -\Gamma/2$. In the remainder of this paper we will look at external heating rates of up to $S_E=0.2 S_{E0}$ that are typical for macroscopic ion traps.

For the evaluation of the second-order Doppler shift due to motion along one spatial direction, it is not the total kinetic energy in a certain mode that is of interest, but rather the total kinetic energy in the secular motion of the clock/spectroscopy ion. This energy is given by the sum of the energies in both modes along that spatial axis, weighed by the relative fraction of clock ion energy to the total energy in the respective mode:
\begin{equation}
E_{c} = \left( b_2^2 E_i + b_1^2 E_o\right).
\end{equation}

\begin{figure}
\includegraphics{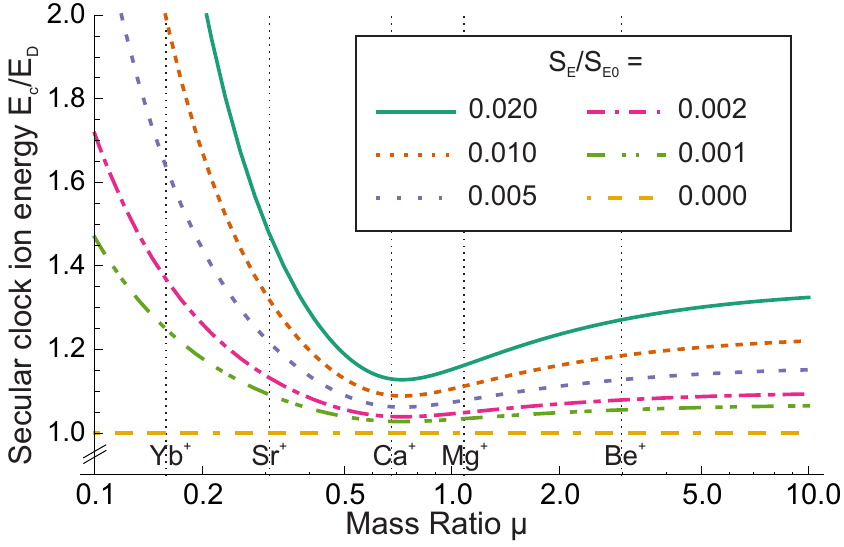}
\caption{Normalized axial clock ion energy plotted against the mass ratio $\mu$ of the ion crystal. The energy $E_c$ of the clock ion is the sum of the clock ion energy in both modes. The calculations were performed for varying electric field spectral densities $S_E$. The cooling laser intensity $I/I_0$ has been optimized for each value of $S_E$. The dashed lines show different logic ion species for an Al$^+$ clock. \label{CoolingLimitsHeating}}
\end{figure}

Figure \ref{CoolingLimitsHeating} shows a plot of $E_{c}$ as a function of the mass ratio $\mu$ for axial motion. The $y$-axis is normalised to the energy of the clock ion at the Doppler-cooling limit without external heating and the electric field spectral densities $S_E$ are given as multiples of $S_{E0}$. In this figure, the intensity was optimised for each data point to achieve the lowest energy in the clock ion. The minimal clock ion energy for different electric field heating rates is always achieved at a mass ratio of $\mu = 8/11$. This plot is independent of $\Gamma$ and the actual values of $m_1,m_2$. However, the normalization factor and therefore the absolute value for the cooling limit depends on the linewidth of the cooling ion. This doubles all cooling limits for Mg$^+$ with respect to all other ions, owing to its twice as large cooling transition linewidth compared to the other ions. The figure shows that the axial clock ion kinetic energy is only a weak function of the mass ratio. This is a direct consequence of the weak dependence of the axial mode amplitudes on this parameter, as shown in Fig.~\ref{FrequenciesTwoIons}.

\begin{figure*}
\includegraphics{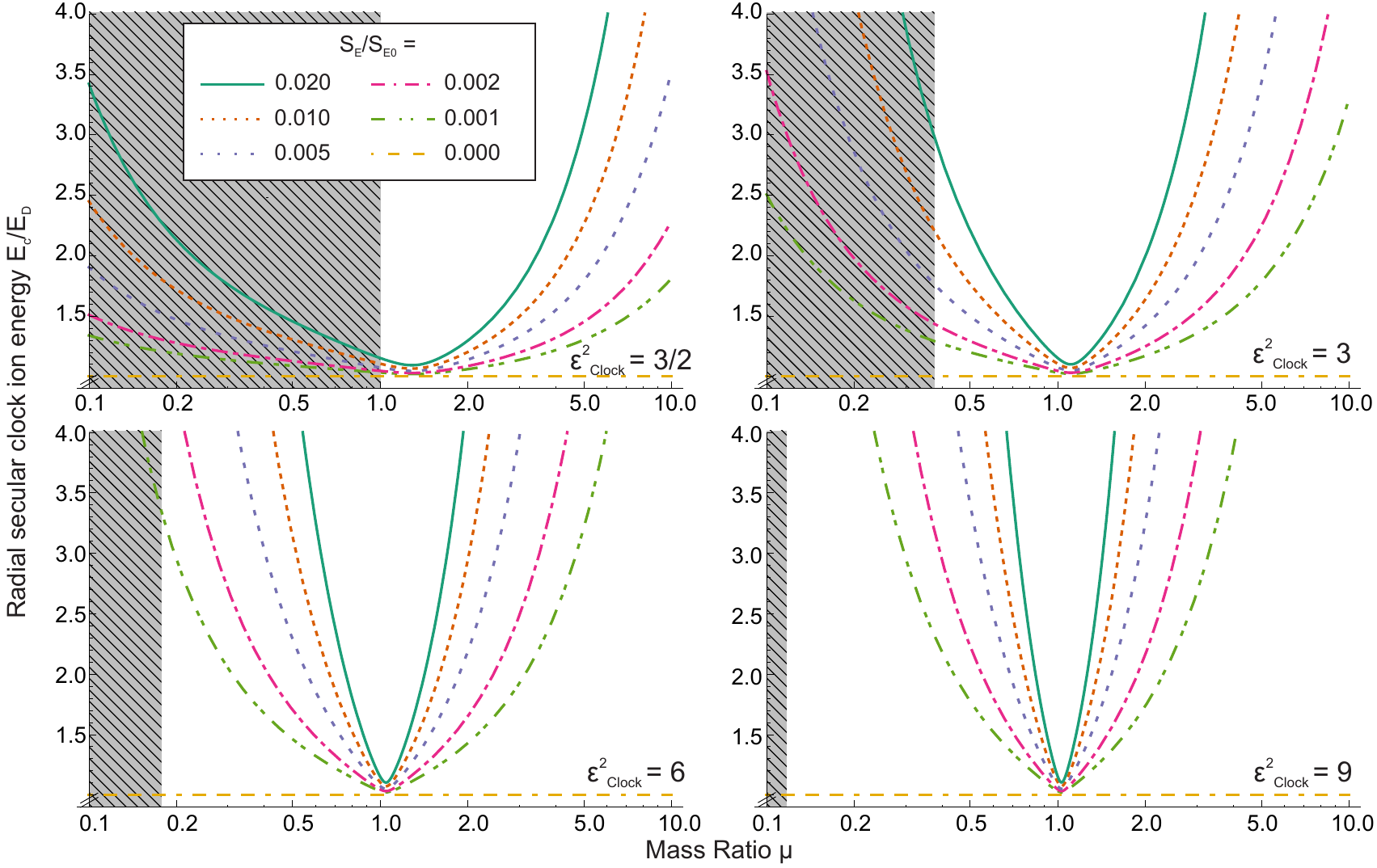}
\caption{Normalized radial clock ion energy plotted against the mass ratio $\mu$ of the ion crystal for radial modes. The energy $E_c$ of the clock ion is the sum of the clock ion energy in both radial modes. The shaded region indicates parameters for which the ion crystal is no longer linear in axial direction. The cooling laser intensity $I/I_0$ been optimized for each value of $S_E$.\label{CoolingLimitsHeatingRadial}}
\end{figure*}

Figure \ref{CoolingLimitsHeatingRadial} shows the normalized clock ion energies in one of the radial directions for varying mass ratios and different $\epsilon$ parameters, assuming a radially symmetric trap ($\alpha = 1/2$). The stated $\epsilon$ is always that of a single clock ion in the trap. The shaded regions in the graphs show the areas in which the crystal is not stable (see \ref{CoolingLimits}) and hence can be ignored. The graphs show that mass ratios slightly larger than 1 always result in the lowest possible clock ion kinetic energy. For a given mass ratio, the energy of the clock ion in the presence of external heating is lowest when operating the trap close to instability of the linear axial configuration. The reason for this is that the larger the radial confinement, the more the modal amplitude in the radial modes tend towards $b_1 \approx 1$, $b_2 \approx 0$ or vice versa (see Eq.~(\ref{RadialBFactor}) and Fig.~\ref{FrequenciesTwoIons}). In that case, the mode with negligible motion of the cooling ion will cool very poorly, leading to an elevated steady state temperature in the presence of external fields. In this regime, the comparison of Fig. \ref{CoolingLimitsHeating} and \ref{CoolingLimitsHeatingRadial} (the y-axes are normalised to the same energies) shows that the heating-induced radial clock ion energy is larger than the axial clock ion energy and will therefore dominate the clock frequency shifts.

Choosing an $\epsilon$ parameter close to the instability regime improves the cooling limits, but at the same time increases the intrinsic micromotion amplitude of the radial modes (see Appendix \ref{AppendixMicromotion}). Fig.~\ref{MicroMotionFactorTotalEnergy} shows the normalized total kinetic energy (equal to the sum of secular and micromotion kinetic energy) of the clock ion in an ion crystal that has been cooled to the Doppler cooling limit (i.e. without external heating) plotted against the crystal mass ratio.

\begin{figure}
\includegraphics{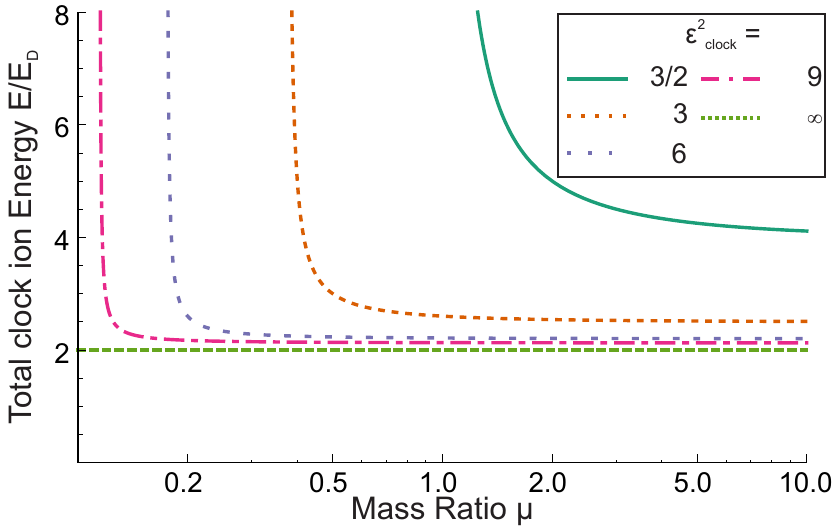}
\caption{Total clock ion energy (secular energy + micromotion energy) of a clock ion in a Doppler cooled crystal along one radial direction. Here the mass ratio $\mu$ and the $\epsilon$-parameters were varied and the energy was normalised to the Doppler cooling energy $E_\mathrm{D}$. The absence of external heating was assumed. Due to large micromotion contributions the total energy diverges at the points where the crystal becomes instable.\label{MicroMotionFactorTotalEnergy}}
\end{figure}

Since the micromotion kinetic energy contributes to the second-order Doppler shift in the same way the secular energy does, it is not advisable to perform spectroscopy in a trap operating close to the unstable regime. The operation in a trap with very strong radial confinement is equally bad because of the inefficient cooling of the weakly damped radial mode. For best cooling performance it is therefore advisable to optimise the $\epsilon$-parameter according to the observed heating rate in the trap.

\begin{figure*}
\includegraphics{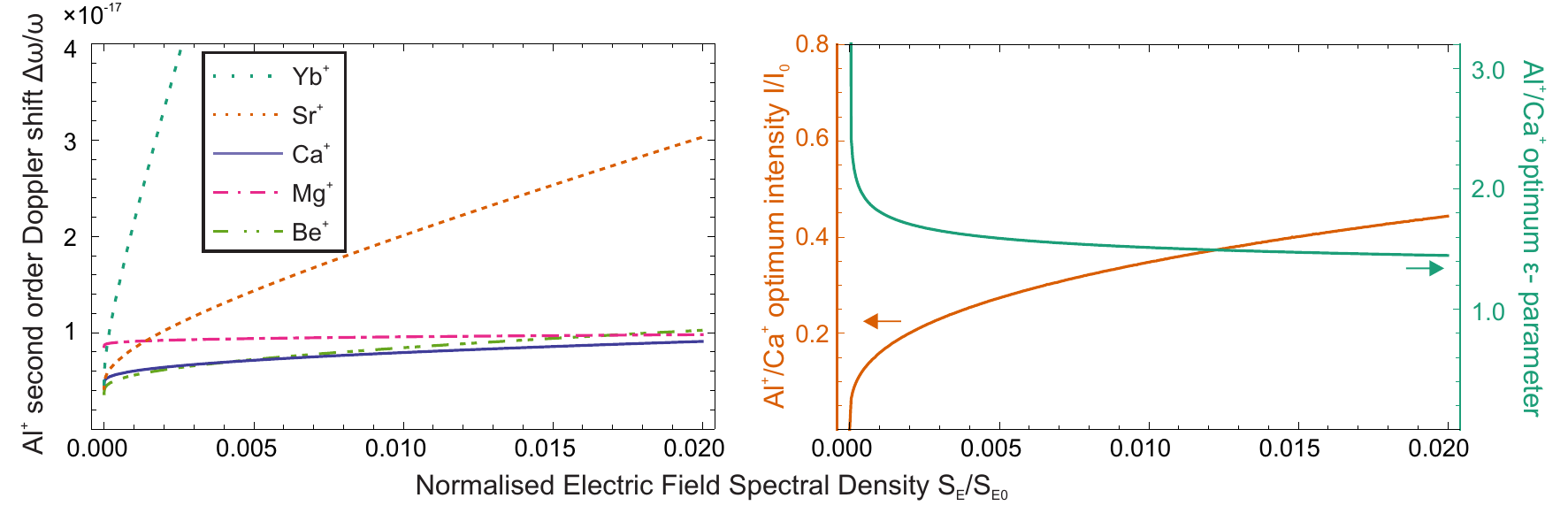}
\caption{Comparison of secular motion-induced second-order Doppler shifts for Al$^+$/X$^+$-crystals with different cooling ions X$^+ =$\{Be$^+$, Mg$^+$, Ca$^+$, Sr$^+$, Yb$^+$\} in the presence of external heating with an electric field spectral density of $S_E$. The right Figure shows the parameters $\epsilon$ and $I/I_0$ for best cooling performance for an Al$^+$/Ca$^+$-crystal. Those optimum parmaters for a spectral energy density of $S=0.005 S_{E0}$ are given by ${2.63,0.40}$,${3.17,0.05}$,${1.12,0.87}$ and ${1.04,1.61}$ for the Be, Mg, Sr and Yb crystals, respectively.\label{CoolingLimitsHeatingTotalShift}}
\end{figure*}

A comparison of the total second-order Doppler shift (including the scaling factors due to micromotion for the radial modes) in a linear ion trap for Al$^+$/X$^+$-ion crystals with different possible cooling ions X$^+$ is shown in Fig. \ref{CoolingLimitsHeatingTotalShift}. The $y$-axis gives the normalised electric field spectral density $S_E/S_{E0,\mathrm{Ca}}$ \cite{SE0_calcium}. A value of $S_E/S_{E0}=0.02$ corresponds to a radial heating rate of a single Ca$^+$ ion in a trap with radial trap frequencies of 2.5 MHz of roughly 1500 quanta per second and can be regarded as an upper limit for most ion traps used for spectroscopy. The $\epsilon^2$ value as well as the Doppler laser detuning $\Delta$ and intensity $I/I_0$ were optimised for each value of $S_E$.

The graph shows that an Al$^+$/Mg$^+$ crystal suffers the least from external heating since its mass ratio is very close to 1. However, for traps with fairly low heating rates, the Doppler shift in Al$^+$/Ca$^+$ and Al$^+$/Be$^+$ traps will be lower than that of Mg$^+$-systems because the linewidth of the cooling transition of Mg$^+$-ions is approximately a factor of 2 larger than that of Ca$^+$ and Be$^+$-ions. The two other species Yb$^+$ and Sr$^+$ suffer from their small mass ratios and require traps with very small heating rates to reach comparable performance.

At this point it should be noted, that not the absolute value of the Doppler shift, but rather its uncertainty is relevant for the clock performance. However, since this uncertainty can be assumed to scale with the absolute shift, a reduction of the latter will result in a reduction of the former and therefore lead to better clock performance. If Doppler cooling is turned off during spectroscopy, careful modelling or measurement of the heating process is required to establish the uncertainty.

\section{Cooling Times\label{CoolingTimes}}
An atomic clock will reach it's maximum performance in terms of stability, if the clock transition is probed without any dead time between consecutive readings~\cite{dick_local_1988, santarelli_frequency_1998, peik_laser_2006}. One contribution to the dead time is the time it takes to cool the ion crystal in case a collision with a hot background gas particle occurred. The large energy transfer to the ion crystal leads to typical temperatures on the order of room temperature, resulting in the decrystallization of the ion crystal. As a consequence, it is necessary to take the non-linear contributions of the Coulomb forces into account to evaluate the cooling dynamics at these high temperatures. The nonlinearities are usually neglected in the small oscillation approximation used to describe the low temperature dynamics of the system that we have used in the previous sections. The dynamics of small ion crystals during laser cooling in various temperature regimes have first been studied in the context of phase-transitions and chaos theory \cite{bluemel_phase_1988, bluemel_chaos_1989}. 

Here, we numerically solved the equations of motion resulting from the complete ponderomotive two ion potential 
\begin{multline}
V(x_1,y_1,z_1,x_2,y_2,z_2)= \frac{1}{2}m_1\left( \omega_{x1}^2 x_1^2 + \omega_{y1}^2 y_1^2 + \omega_{z1}^2 z_1^2\right) \\
+ \frac{1}{2}m_2\left( \omega_{x2}^2 x_2^2 + \omega_{y2}^2 y_2^2 + \omega_{z2}^2 z_2^2\right) \\
+ \frac{e^2}{4 \pi \epsilon_0} \frac{1}{\sqrt{(x_1-x_2)^2+(y_1-y_2)^2+(z_1-z_2)^2}} ,
\end{multline}
where $\omega_{kj}$ is the trap frequency of ion $j$ in direction $k\in{x,y,z}$. and $k_j$ are the position coordinates.
For the simulations, the RF potential was neglected. However, we expect only minor modifications to the results presented here when micromotion is included in the treatment \cite{marciante_ion_2010}.
We used an adaptive Runge-Kutta method where the step size was reduced significantly every time the ions got close enough to explore the high non-linearities of the $1/r$-Coulomb potential. Cooling was incorporated into the calculation by multiplying the instantaneous scattering rate $R$ at every time step with the instantaneous step size $dt$ at the same step and comparing the resulting value with a random number $rn(0,1)$ between 0 and 1 from a "Mersenne Twister"-type pseudo random number generator \cite{Matsumoto1998}. A scattering event that changed the velocity $\mathbf{v}_1$ of the cooling ion according to the momentum change due to both absorption and spontaneous emission was therefore calculated whenever the condition
\begin{equation}
R \, dt = \frac{\Gamma}{2} \frac{I/I_0}{1 + I/I_0 + \left(2 (\Delta - \bm{k} \cdot \bm{v}_1) / \Gamma \right)^2} dt < rn(0,1). \label{ScatteringRate}
\end{equation}
was fulfilled.

With a small enough step size the simulation turned out to be very robust and reproduced two-ion cooling limits as well as theoretical curves for the cooling duration of single ions (e.g. \cite{Wesenberg2007}). 

\begin{figure*}
\includegraphics{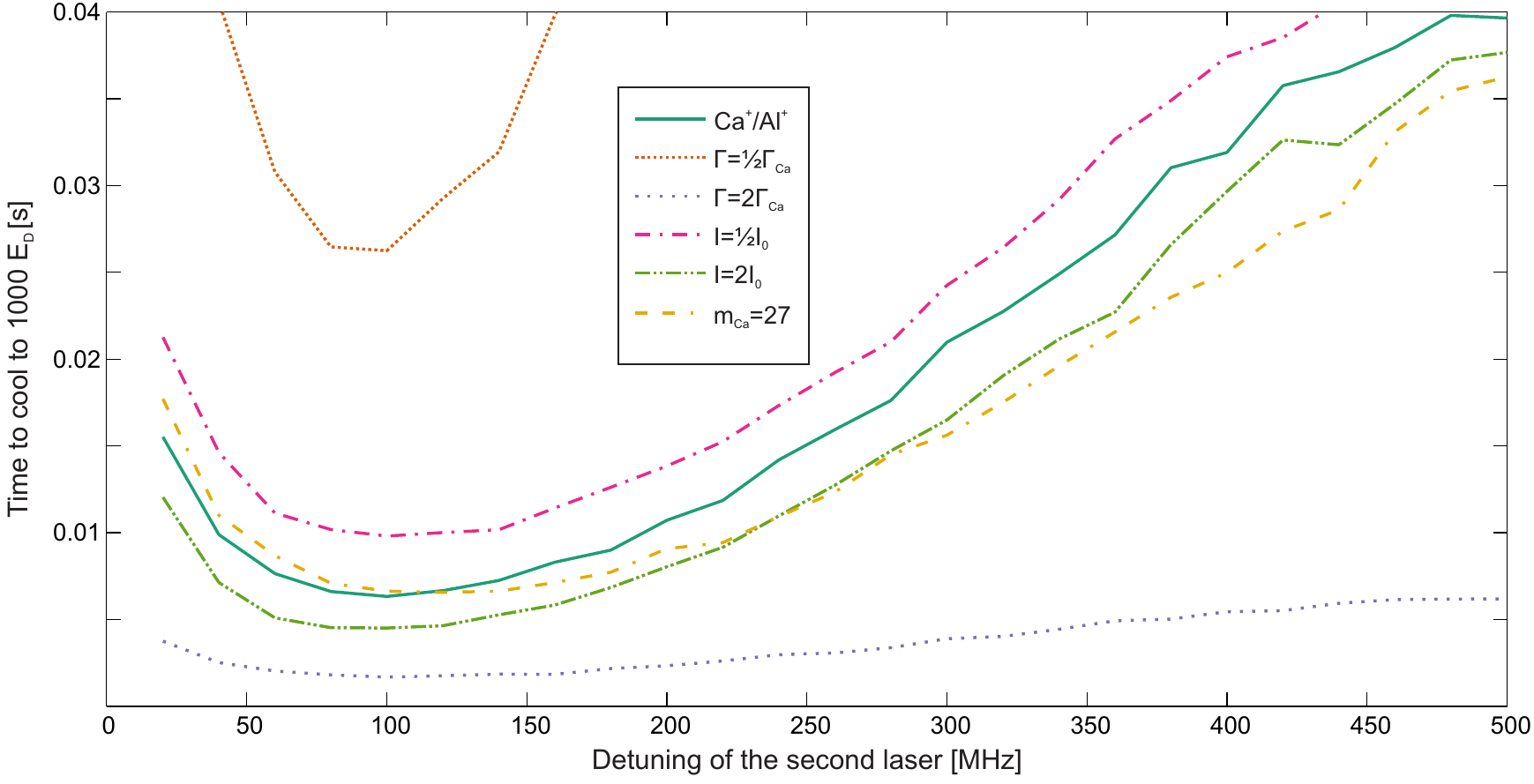}
\caption{Time needed to cool an Al$^+$/Ca$^+$-ion pair from 19.3~K to 1000 Doppler limits. The $y$-axis shows the detuning of a second laser at saturation intensity that is cooling the ion crystal in addition to the standard $-\Gamma/2$-detuned beam.\label{TwoIonsCoolingTo1000DopplerLimits}}
\end{figure*}

A typical collision event with the background gas will be an elastic collision of a hydrogen molecule with either one of the two cold ions in the crystal. The maximum energy that can be transmitted in an elastic collision of two particles with masses $m_1, m_2$ if one of them is initially at rest is given by $\frac{4 m_1 m_2}{(m_1+m_2)^2}$ times the kinetic energy of the hot particle. In case of the collision of a cold aluminium ion and a hydrogen molecule at 300~K this means that a maximum energy of $\approx 0.26 \times 3/2 k_B \times 300$~K can be transferred to the aluminium ion. Here $k_B$ denotes the Boltzmann constant. This corresponds to a temperature of the ion crystal of $\approx 19.3$~K. Cooling an ion crystal from these high temperatures is aided by adding a far detuned laser beam additionally to the standard cooling beam at $\Delta = -\Gamma/2$. This enhances the scattering rate of the rapidly moving cooling ion with large Doppler detuning. The optimum detuning of the second laser was found by simulating the cooling time for an aluminium/calcium ion pair starting at a temperature of 19.3~K as a function of the detuning. The results are shown in Fig.~\ref{TwoIonsCoolingTo1000DopplerLimits}. 

Both lasers were assumed to be directed onto the ion from the $(1,1,1)$ direction and both had one saturation intensity at the position of the ion. To compare the performance of the different detunings, the time to reach a crystal energy of 1000 Doppler cooling limits was simulated. This energy roughly corresponds to the energy at which the ions crystallize. Fastest cooling was achieved at a detuning of $\sim$-100~MHz. This optimum detuning of the second laser beam proved to be very robust against changes in the cooling parameters, such as a change in linewidth of the cooling transition, the intensity of the cooling laser, or the mass of the cooling ion. Since these changes mostly affect the scattering rate, the times needed to cool vary strongly. The general characteristics of the curve with the minimum close to -100~MHz, however, stays the same. This can be explained by the fact that in a well-thermalized regime, the cooling ion will carry half the crystal energy and the optimal detuning should only depend on the average cooling ion velocity in the direction of the incoming laser. This is independent of the transition linewidth or the intensity of the cooling light. The change of mass of the cooling ion to smaller values shifts the optimal detuning to slightly higher values. This is because a lighter cooling ion has higher average velocities if it has the same energy as a heavier ion.

\begin{figure*}
\includegraphics{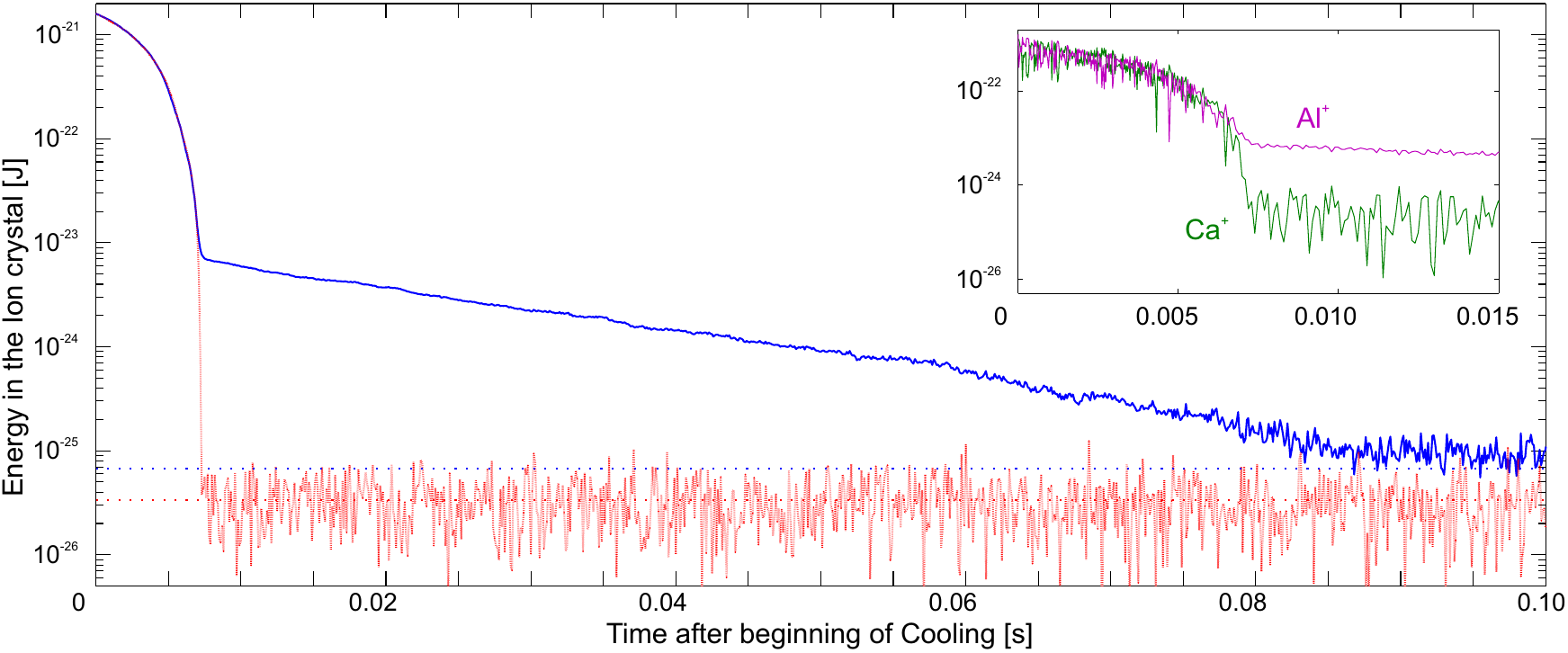}
\caption{Cooling evolution of an Al$^+$/Ca$^+$ ion pair after a collision event with an H$_2$-molecule (blue line). The red line shows the cooling evolution of a single Ca$^+$ ion in the same trap with the same initial energy. The horizontal dotted lines denote the Doppler cooling limit and the inset shows the energies of the calcium and aluminium ion before and after crystallisation occured at around 7~ms.\label{TwoIonsComparedToOneIon}}
\end{figure*}

The difference in cooling dynamics between a two- and a single-ion system is shown in Fig.~\ref{TwoIonsComparedToOneIon}. Besides the Al$^+$/Ca$^+$ pair, the results for a single Ca$^+$ ion in the trap with the second laser detuned by 140~MHz is shown. This slightly higher optimum detuning arises, since a single ion will on average have higher velocities than the cooling ion in a two-ion system since the latter can exchange energy with the clock ion. For the simulation, the starting energy of the single Ca$^+$ ion was set to the same value as that of the two-ion pair, although a collision with a hydrogen molecule would lead to a lower initial temperature.

The cooling rates match extraordinarily well at the beginning of the cooling process. This is because in the high temperature regime the two ions in the crystal collide very often so that all motional modes are thermalized almost instantly. Therefore no poorly cooled modes exist that would decrease the cooling rates. However, as soon as the ion pair crystallises (around 7~ms), no collisions thermalize the modes and the cooling rate decreases significantly owing to the weakly damped radial modes (see inset in Figure~\ref{TwoIonsComparedToOneIon}). The crystallisation effect can be visualised by plotting the energy of the Ca$^+$ and Al$^+$ separately over the cooling time, as shown in the inset of Figure \ref{TwoIonsComparedToOneIon}. Once crystallisation occurs, the energy exchange between the motional modes of the two ions is prohibited and the cooling rate is determined by the mode amplitudes of the cooling ion (for a theoretical description of the crystallisation see for example \cite{bluemel_chaos_1989} for a measurement of the damping of the weakly damped modes see the supplementary material of \cite{Rosenband2008}).

Summarizing the simulation results of Fig.~\ref{TwoIonsCoolingTo1000DopplerLimits}~and~\ref{TwoIonsComparedToOneIon}, an optimum cooling strategy after a background gas collision can be derived: The fastest cooling is achieved by ramping the laser detuning according to the actual energy of the ion crystal, such that maximal scattering rates are maintained. Additionally, the intensity of the cooling laser should be large to increase the rates even further. Once crystallization occurs, the laser should be ramped to a detuning and an intensity for which the lowest energies are achieved (typically $\Delta \approx \Gamma/2$, $I\sim I_0$).

In our simulations the time needed to cool an Al$^+$/Ca$^+$-crystal to 2 Doppler-Cooling limits is roughly 90~ms (which corresponds well to previous studies \cite{Schiller2003,Drewsen2003}) while the time it takes to reach crystallisation is only 7~ms. A possibility to reduce the final cooling time is to couple the well-damped to the poorly-damped modes by a static radial electric field as has been demonstrated in \cite{Rosenband2008}. Even without such a scheme, a Doppler cooling time of $<100$~ms corresponds to less then one measurement cycle in state of the art quantum logic clocks (i.e. cycle time 230~ms \cite{Chou2010}) and since such a random collision should only occur roughly every 100~s in a good vacuum environment, its contribution to the total dead time will be insignificant compared to dead times due to detection and state preparation. However, it seems necessary that the collision events are detected and actions are taken to ensure fast recrystallization, since the typical Doppler cooling time used in single ion experiments are on the order of ms, which will not be sufficient to cool the crystal to its cooling limit.

\section{Summary\label{Summary}}
In this paper we examined the steady state sympathetic cooling limits with and without external heating of different two-ion two-species ion crystals with an emphasis on second-order Doppler shifts that pose a limit to precision spectroscopy. We find that in the absence of external heating the Doppler cooling limit can be reached in all six normal modes of the ion crystal, independent of the mass ratio between the two ions. However, with additional external heating, the cooling limit becomes a sensitive function of the normal mode amplitudes. The modes where the cooling ion has a large amplitude are cooled most efficiently and reach a low cooling temperature. We find that operating the trap close to instability with respect to a flip of the linear-axial to the linear-radial orientation enhances the amplitudes of the critical radial modes. This effect has to be balanced with the increased micromotion in this regime. Specifically, we find that three logic ion species candidates, ${}^{25}$Mg$^+$, ${}^{9}$Be$^+$ and ${}^{40}$Ca$^+$, are most suitable as sympathetic cooling ions for an aluminium ion clock, even in the presence of moderate external heating rates. The former is a good choice because its mass ratio relative to aluminium is close to one which means that no poorly-damped modes exist that would make the crystal vulnerable to high external heating rates. The latter two have a cooling transition with a smaller linewidth than the ${}^{25}$Mg$^+$-transition resulting in lower Doppler-limits and therefore lower second-order Doppler shifts. For traps with very low heating rates, even heavier sympathetic cooling ions, such as ${}^{88}$Sr$^+$ will perform well. Furthermore, we investigated the cooling time of the ${}^{40}$Ca$^+$/${}^{27}$Al$^+$ unequal mass ion crystal after a collision with background gas. Monte-Carlo cooling simulations taking into account the ponderomotive trap potentials and the Coulomb-potentials revealed cooling times of roughly 100~ms, which would not pose a limitation for the clock stability. However, these events should be detected immediately and counteracted by applying a second (further detuned) laser beam for rapid recrystallization. We therefore believe that an ${}^{40}$Ca$^+$/${}^{27}$Al$^+$ quantum logic clock will lead to a similar clock performance as ${}^{25}$Mg$^+$/${}^{27}$Al$^+$ clocks and might even outperform them in terms of second-order Doppler shifts for ion-traps with low heating rates. Furthermore, the presented results are relevant for high precision spectroscopy of other sympathetically cooled ion species, such as highly-charged or molecular ions.

\appendix
\section{Derivation of the Cooling and Heating Rates} \label{RatesCalculation}
The cooling rate of a two-ion crystal mode can be calculated by determining the energy change in this mode for every scattering event. According to Eq. (\ref{EnergyI}) the energy change in an in-phase mode (analogously for out-of-phase modes) with modal amplitudes $z_i$ before and $z_i^\prime$ after the absorption of a cooling photon is given by
\begin{equation}
 \Delta E_i = \frac{1}{2}m_1 \omega_i^2 \left( z_i^{\prime 2} - z_i^2\right). \label{DeltaEi}
\end{equation}
$z_i^\prime$ can be calculated by expressing the modal amplitude $z_i$ as a function of the positions $q_1,q_2$ and velocities $v_1 = \dot q_1, v_2 = \dot q_2$ along the axis of the relevant modes of the two ions before the absorption
\begin{equation}
 z_i = \sqrt{\frac{(\sqrt{\mu} b_2 v_2 + b_1 v_1)^2+\omega_i^2 (\sqrt{\mu} b_2 z_2 + b_1 z_1)^2}{\omega_i^2}}
\end{equation}
and then adding the velocity change $\mathrm{d}v = l_x \hbar k/m_1$ (exemplarily for the $x$-direction) to the cooling ion velocity
\begin{equation}
 z_i^\prime = \sqrt{\frac{(\sqrt{\mu} b_2 v_2 + b_1 (v_1+\mathrm{d}v))^2+\omega_i^2 (\sqrt{\mu} b_2 z_2 + b_1 z_1)^2}{\omega_i^2}}.
\end{equation}
Substituting these equations into Eq. (\ref{DeltaEi}) $\Delta E_i$ gives
\begin{align}
 \Delta E_i &= \frac{1}{2} m_1 b_1^2 \mathrm{d}v^2 +  m_1 b_1 \mathrm{d}v (b_1 v_1 + \sqrt{\mu} b_2 v_2) \\
&= \frac{1}{2} m_1 b_1^2 \mathrm{d}v^2 + m_1 b_1 \mathrm{d}v \omega_i z_i \cos{\left(\omega_i t + \phi_i \right)}. 
\end{align}
The first term of this equation is a constant heating that is taken care of in the heating rate. The second term corresponds to mode cooling if $\cos{\left(\omega_i t + \phi_i \right)}$ is negative and heating if it is positive. The rate $R$ at which scattering events occur is given by Eq. (\ref{ScatteringRate}). The cooling rate is computed by averaging the product $R \times (\Delta E_i-\frac{1}{2} m_1 b_1^2 \mathrm{d}v^2)$ over the oscillation periods of all six crystal modes
\begin{multline}
 \frac{dE_i}{dt}_{\text{Cooling}} = \\
 \frac{1}{(2\pi)^6}\overset{2\pi}{\underset{0}{\int \!\!\! \int \!\!\! \int \!\!\! \int \!\!\! \int \!\!\! \int}} R (\Delta E_i-\frac{1}{2} m_1 b_1^2 \mathrm{d}v^2) d^3\phi_{i} \, d^3\phi_{o}.
\end{multline}
The integration over all six modes is necessary because the scattering rate depends on the product $\bm{k} \cdot \bm{v_1}$ and therefore on the cooling ion velocity in all three spatial dimensions. This integral cannot be solved in general but for cold crystals, where $|kv_1|\ll \Gamma$ in all three dimensions, the integrand can be expanded and higher order terms in $k v_1 / \Gamma$ can be neglected, leading to Eq. (\ref{CoolingRateEquation}).

The heating rate computes similarly by averaging the product $R \times \frac{1}{2} m_1 b_1^2 \mathrm{d}v^2$ over the six modes which leads in first order to a velocity independent term
\begin{equation}
 \hbar^2 3 l_x^2 k^2 \frac{I}{I_0} \frac{\Gamma}{12 m_1} \frac{b_1^2}{1+I/I_0 + (2\Delta/\Gamma)^2}. \label{HeatingContributionAbsorption}
\end{equation}
To this heating due to the absorption of the photons one has to add a contribution of the spontaneous emission. This is exactly as large as the absorption effect but does not depend on the direction of the cooling laser. Assuming an isotropic emission it is given by 
\begin{equation}
 \hbar^2 k^2 \frac{I}{I_0} \frac{\Gamma}{12 m_1} \frac{b_1^2}{1+I/I_0 + (2\Delta/\Gamma)^2} \label{HeatingContributionEmission}
\end{equation}
so that the sum of Eq. (\ref{HeatingContributionAbsorption}) and (\ref{HeatingContributionEmission}) gives the total heating rate as in Eq. (\ref{TotalHeatingRate}).

\section{Doppler Shift Contribution of Intrinsic Micromotion}\label{AppendixMicromotion}
To calculate the amplitude of the micromotion in the radial modes of the two-ion crystal, we generalize the derivation for single ions given in \cite{Berkeland1998}. The force on a single ion in radial direction (we exemplarily us the $x$-direction in the following) is given by
\begin{equation}
 F_x = -\partial_x e \Phi(x,y,z,t) = 2 x \alpha \frac{e U_0}{d^2} - x \frac{e V_0}{R^2} \cos \Omega_t t
\end{equation}
where $\Phi(x,y,z,t)$ from Eq. (\ref{TotalPotential}) was used. In two-ion crystals an additional force arises due to the Coulomb potential $U_C$ between the two ions
\begin{multline}
 U_C(x_1,x_2,y_1,y_2,z_1,z_2) = \\
\frac{e^2}{4 \pi \epsilon_0} \frac{1}{\sqrt{(x_1-x_2)^2+(y_1-y_2)^2+(z_1-z_2)^2}}.
\end{multline}
The Coulomb-force in $x$-direction is given by
\begin{multline}
 F_{C,x} = -\partial_x  U_C(x_1,x_2,y_1,y_2,z_1,z_2) = \\
\frac{e^2}{4 \pi \epsilon_0} \frac{1}{\sqrt{(x_1-x_2)^2+(y_1-y_2)^2+(z_1-z_2)^2}^3} (x_1-x_2). \label{CoulombForce1}
\end{multline}
We now assume that the ion crystal is cold enough so that the ions oscillate with small amplitudes around their equilibrium positions $\bm{x_1^{(0)}} = (0,0,z_0),\, \bm{x_2^{(0)}} = (0,0,-z_0)$ where $z_0 = \left( \frac{e d^2}{32 U_0 \pi \epsilon_0}\right)^{1/3}$ (see for example \cite{james_quantum_1998}). Linearising of Eq. (\ref{CoulombForce1}) in $(x_1-x_2)$ around $\bm{x_1^{(0)}}, \bm{x_2^{(0)}}$ gives
\begin{equation}
 F_{C,x} \approx \frac{e^2}{4 \pi \epsilon_0} \frac{1}{\sqrt{(2 z_0)^2}^3} (x_1-x_2) = \frac{e U_0}{d^2} (x_1-x_2). \label{CoulombForce2}
\end{equation}
 If we now consider only a single mode (i.e. the in-phase or the out-of-phase mode), we can write the $x$-motion of the second ion as a constant factor $\beta$ times the motion of the first ion
\begin{equation}
 x_2 = \beta x_1
\end{equation}
where $\beta = b_2/(\sqrt{\mu}b_1)$ for in-phase and $\beta = -b_1/(\sqrt{\mu}b_2)$ for out-of-phase modes as follows from Eq. (\ref{IonMotion1}) and (\ref{IonMotion2}).

Replacing $x_2$ in Eq. (\ref{CoulombForce2}) and adding the Coulomb-force to the force due to trap potentials, the total force $G_{x}$ on the ion is given by
\begin{equation}
 G_x = F_x + F_{C,x} =  x_1 \frac{e U_0}{d^2} (2 \alpha + 1 - \beta) - x_1 \frac{e V_0}{R^2} \cos \Omega_t t.
\end{equation}
Introducing the $a_x = - \frac{8 \alpha e U_0}{m_1 d^2 \Omega_T^2}$ and $q_x = \frac{2 e V_0}{m_1 R^2 \Omega^2}$ as in \cite{Berkeland1998}, the equation of motion for $x_1$ can be given in typical Mathieu form
\begin{equation}
 \ddot x_1 + \left( a_x \left( 1 +\frac{1-\beta}{2\alpha} \right) + 2 q_x \cos \left(\Omega_T t \right) \right) \frac{\Omega_T^2}{4} x_1 = 0.
\end{equation}
Compared to the equation of motion for a single ion ($\ddot x_1 + \left( a_x + 2 q_x \cos \left(\Omega_T t \right) \right) \frac{\Omega_T^2}{4} x_1 = 0$), the only difference is that the $a_x$ is varied by a factor $(1+ (1-\beta)/(2\alpha))$ in the case of a two-ion crystal mode. Accordingly the first order solution \cite{Landau1976} (valid for $q_x \ll 1, \, a_x (1+ (1-\beta)/(2\alpha)) \ll 1$) of the single ion equation is still valid, if that factor is included:
\begin{equation}
 x_1(t) \approx x_i b_1 \sin{\omega_{x,i} t + \phi_{x,i}} \left(1 + \frac{q_x}{2} \cos(\Omega_T t) \right)
\end{equation}
In this equation the amplitude $x_i b_1$ of the oscillation was already chosen to comply with Eq. (\ref{IonMotion1}). $\omega_{x,i}$ denotes the in-phase mode frequency and is given by 
\begin{equation}
 \omega_{x,i} = \frac{1}{2} \Omega_T \sqrt{a_x (1+ (1-\beta)/(2\alpha)) + \frac{1}{2}q_x^2}.
\end{equation}
The average squared velocity of this motion is given by
\begin{align}
 < \dot x_1^2 > &= \frac{1}{2} (x_i b_1)^2 \left( \omega_{x,i}^2 + \frac{1}{8} q_x^2 \Omega^2 \right) \notag \\
&= \frac{1}{2} (x_i b_1)^2 \omega_{x,i}^2 \left( 1 + \frac{q_x^2}{q_x^2 + 2 a_x (1+ \frac{1-\beta}{2\alpha})} \right) \notag \\
&= \frac{1}{2} (x_i b_1)^2 \omega_{x,i}^2 \left( 1 + \frac{2 \epsilon^2}{2 \epsilon^2 - 2 \alpha - (1- \beta)}\right) \label{MicromotionContribution}
\end{align}
where in the last step the $q_x$ and $a_x$ were substituted by the $\epsilon^2 = \frac{-\alpha q_x^2}{2 a_x}$ factor that is used throughout this paper. Equation \ref{MicromotionContribution} shows that for the radial modes the average squared velocity is not just given by the secular motion but that a micromotion term of the same order of magnitude has to be considered as well. For strong radial confinement $\epsilon^2 \gg 1$, the micromotion contribution approximately equals the contribution from the secular motion. However, for weak confinement, it can become much larger. 

\begin{figure*}
  \includegraphics{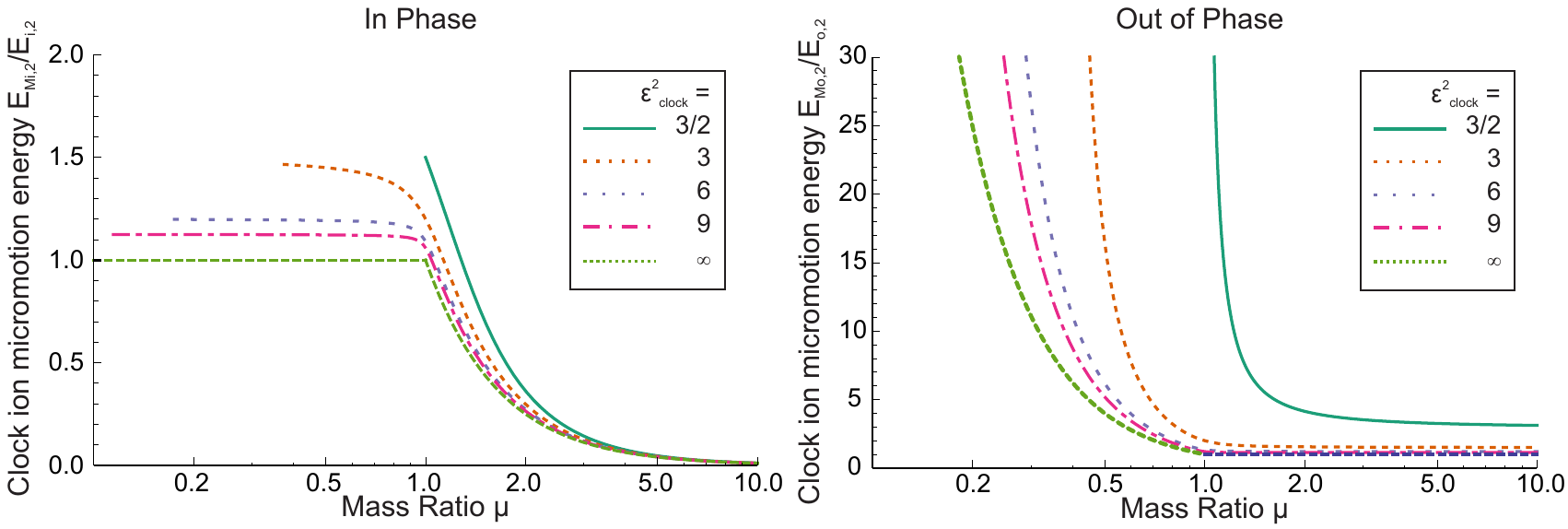}
 \caption{Relative micromotion energy of the clock ion in the in-phase and out-of-phase radial modes. The energy is normalised to the secular energy of the clock ion in the respective mode. In this graph the absence of external heating was assumed. The dashed lines denote the asymptotic behaviour for very large radial confinement ($\epsilon \rightarrow \infty$).}
 \label{MicroMotionFactors}
\end{figure*}

Figure \ref{MicroMotionFactors} shows a plot of the relative kinetic energy of the micromotion of the clock ion compared to its secular energy for one pair of radial modes and for different $\epsilon$-parameters (again the stated $\epsilon$ is that of a single clock ion) in the absence of external heating. The graph shows that the out-of-phase fractional micromotion energy contribution can get very large, when the trap is operated close to the instability limit.

The micromotion contribution factor for the second ion (which is in this paper the clock ion and therefore the interesting one), can be calculated identically but the $\epsilon_2 = \sqrt{m_1/m_2} \epsilon$ and $\beta_2 = 1/\beta$ parameters for the second ion have to be used. Expressed in terms of the $\epsilon$ and $b_1, b_2$ parameters, the average squared velocity of the clock ion is therefore given by
\begin{align}
  < \dot x_{2,i}^2 > &= \frac{1}{2 \mu} (x_i b_2)^2 \omega_{x,i}^2 \!\! \left( \!\! 1 \!\! + \!\! \frac{2 \epsilon^2/\mu}{2 \epsilon^2/\mu \!\! - \!\! 2 \alpha \!\! - \!\! ( \! 1 \!\! - \!\! \sqrt{\mu}b_1/b_2)}\right) \\
  < \dot x_{2,o}^2 > &= \frac{1}{2 \mu} (x_o b_1)^2 \omega_{x,o}^2 \!\! \left( \!\! 1 \!\! + \!\! \frac{2 \epsilon^2/\mu}{2 \epsilon^2/\mu \!\! - \!\! 2 \alpha \!\! - \!\! ( \! 1 \!\! + \!\! \sqrt{\mu}b_2/b_1)}\right)
\end{align}
for the in-phase and out-of-phase mode, respectively. 


\bibliographystyle{apsrev}
\bibliography{paper01,symp_cooling}

\end{document}